\documentclass[pdflatex,sn-mathphys-num]{sn-jnl}

\usepackage{graphicx}
\usepackage{multirow}
\usepackage{amsmath,amssymb,amsfonts}
\usepackage{amsthm}
\usepackage{mathrsfs}
\usepackage[title]{appendix}
\usepackage{xcolor}
\usepackage{textcomp}
\usepackage{manyfoot}
\usepackage{booktabs}
\usepackage{algorithm}
\usepackage{algorithmicx}
\usepackage{algpseudocode}
\usepackage{listings}
\usepackage{subcaption}
\usepackage{natbib}
\usepackage[most]{tcolorbox}
\usepackage{fancyhdr}
\newtcblisting{promptbox}[1][]{
  listing only,
  breakable,
  colback=gray!5,
  colframe=black,
  left=1mm,right=1mm,top=1mm,bottom=1mm,
  fontupper=\ttfamily\small,
  title={Prompt},
  #1
}

\theoremstyle{thmstyleone}

\theoremstyle{thmstyletwo}

\theoremstyle{thmstylethree}%

\begin{document}

\title[Article Title]{Breaking Bad Financial Habits: How LLM Conversations Correct Financial Misconceptions}

\author*[1,3]{\fnm{Jillian} \sur{Ross}}\email{jillianr@mit.edu}

\author[2]{\fnm{Eric} \sur{So}}\email{eso@mit.edu}
\author[2,3]{\fnm{Andrew W.} \sur{Lo}}\email{alo-admin@mit.edu}

\affil*[1]{\orgdiv{Department of Electrical Engineering \& Computer Science}, \orgname{MIT}, \orgaddress{\city{Cambridge}, \state{MA}, \country{USA}}}

\affil[2]{\orgdiv{Sloan School of Management}, \orgname{MIT}, \orgaddress{\city{Cambridge}, \state{MA}, \country{USA}}}

\affil[3]{\orgdiv{Laboratory of Financial Engineering}, \orgname{MIT}, \orgaddress{\city{Cambridge}, \state{MA}, \country{USA}}}

\abstract{
Financial misconceptions carry direct economic costs, from panic selling to equity market avoidance, yet they are notoriously resistant to correction. Traditional financial literacy interventions are constrained by cost, reach, and a persistent gap between knowledge and behavioral change. Across three pre-registered studies, we find that purposefully designed LLMs can durably correct financial misconceptions. Critically, two factors are necessary for this effect. First, corrective intent: LLMs prompted only to discuss a misconception produce corrections no better than unassisted self-reflection, and undirected LLM conversations can actively entrench misconceptions. Second, recipient receptivity: financial concepts are often foreign to the investors who misapply them, and LLM responses pitched below a participant's financial sophistication are judged as less credible and produce substantially weaker corrections. LLMs thus offer a scalable alternative to traditional financial literacy intervention, but only when designed with both factors in mind.
}


\maketitle

\makeatletter
\def\@oddhead{\hfil\textit{Preprint}\hfil}
\def\@evenhead{\hfil\textit{Preprint}\hfil}
\makeatother

\section{Introduction}
\label{sec:intro}

Behavioral economics has established that the cognitive errors underlying poor financial decisions are remarkably resistant to correction, not merely because people lack information, but because the heuristics driving those errors feel locally valid and are largely impervious to awareness of their presence. Systematic deviations from rational decision-making are not random noise, but instead predictable patterns, rooted in cognitive shortcuts that served adaptive purposes in other contexts, but misfire reliably in financial ones \citep{tversky1974judgment, barberis2003survey}. Crucially, this awareness offers little protection to the individual; warning people about anchoring bias, for instance, does little to prevent them from anchoring \citep{wilson1996new}. These dynamics are particularly consequential for personal financial well-being, where persistent misconceptions translate directly into costly behaviors: buying high and selling low, concentrating risk in single assets, panic-selling during downturns, and avoiding equity markets altogether \citep{barberis2006individual, bailey2011behavioral}. Financial misconceptions and their stubborn persistence compound their damage over decades, making their correction both more economically imperative and, in principle, more tractable, given the availability of verifiable expert knowledge. Yet basic financial concepts are often foreign to the very investors who misapply them, and effective correction must account for what they actually understand.

Correcting such misconceptions at scale has proven difficult. Traditional financial literacy interventions, such as educational programs, informational campaigns, and one-on-one advising, are constrained by cost, reach, and a persistent gap between knowledge acquisition and behavioral change \citep{lusardi2019financial, van2011financial, karakurum2019financial}. Human financial advisors can provide personalized guidance, but are inaccessible to many households due to their cost, and are themselves subject to behavioral biases and conflicts of interest that limit advice quality \citep{linnainmaa2021misguided}. Automated robo-advisors are more scalable, but offer generic recommendations without the interactive capacity to address individual misconceptions through dialogue \citep{d2019promises, hackethal2012financial}. The result is a persistent literacy gap that falls hardest on the populations with the most to gain from long-term equity participation \citep{klapper2015financial}.

Large language models (LLMs) offer a theoretically promising alternative. Unlike static educational content, conversational LLMs can engage individuals interactively, tailor counterarguments to their reasoning, and adapt in real time to the concerns of a particular user. This combination of personalization and scale has no clear precedent in the history of financial education. A growing body of evidence supports the persuasive potential of LLM-mediated dialogue more broadly: conversational LLMs have been shown to shift beliefs through interactive exchanges that tailor counterarguments to individual reasoning \citep{altay2023information, altay2022scaling, aggarwal2023artificial}, and conversational LLMs have been found to outperform incentivized human persuaders in changing minds when provided with basic personal information \citep{salvi2024conversational}. Most relevantly, Costello et al.\ \cite{costello2024durably} demonstrate that brief, tailored LLM conversations can reduce conspiracy beliefs by approximately 20\%, with effects persisting two months later. 

While \citet{costello2024durably} demonstrate that tailored LLM dialogue can shift conspiracy beliefs, financial misconceptions have a different etiology, and thus need to be studied separately. Conspiracy beliefs are identity-expressive, sustained by motivated distrust of institutions, and resistant to falsification by design.  In contrast, financial misconceptions arise from domain-general cognitive heuristics---availability, representativeness, extrapolation from recent experience---that are recognizably wrong once expert reasoning is brought to bear \citep{tversky1974judgment, barberis2003survey}. Because they reflect faulty reasoning rather than motivated belief, financial misconceptions are in principle more tractable targets for evidence-based persuasion, but only if the intervention engages the reasoning process rather than merely supplying information \citep{wilson1996new, lusardi2019financial}. Additionally, financial misconceptions are less obviously problematic than conspiracy theories, and an unprompted LLM may default to sycophantic validation rather than correction. Which mechanisms are necessary and sufficient in the objectively correctable case remains an open question.

We address this question through three pre-registered studies. We propose that two factors are necessary to correct misconceptions with LLMs: \textit{corrective intent},  whether the intervention delivers an explicit corrective intent rather than merely directing attention toward the misconception, and \textit{recipient receptivity},  whether the user perceives the source as credible and its arguments as legible given their level of financial sophistication. In the first study, we establish the baseline corrective effect of LLM-mediated dialogue and document its moderation by chatbot perception. In the second, we directly probe the role of \textit{corrective intent} by comparing an LLM prompted to persuade against three alternatives: an LLM prompted only to engage, a self-articulation condition in which participants elaborate their own reasons for holding the misconception, and an unrelated distractor baseline. In the third, we examine the role of \textit{recipient receptivity} by testing whether tailoring response complexity to a participant's financial sophistication improves corrective efficacy. We find that both factors matter: correction requires not only that the LLM be directed to persuade, but that its arguments be legible and credible to the specific user it is addressing.

\section{Results}
\label{sec:results}

First, we find that LLM-mediated conversations produce large, statistically significant reductions in financial misconceptions that persist at least 10 days after the intervention. This effect is moderated by baseline conviction and chatbot perception: the correction is greatest when misconceptions are strongly held and when participants perceive the chatbot as credible and convincing, with lower financial literacy predicting stronger attributions of credibility. Second, explicit persuasive intent is a necessary condition for misconception correction, and the absence of explicit intent is not merely ineffective but actively harmful: an LLM prompted only to discuss a misconception produces corrections statistically indistinguishable from unassisted self-reflection, while undirected LLM engagement raises entrenchment rates. This suggests that conversational elaboration of a held misconception strengthens it in the absence of a deliberate counter-argument. Third, tailoring response complexity to a participant's financial sophistication improves the efficacy of misconception correction.

\subsection{LLMs durably correct financial misconceptions}
\label{ssec:result1}

Study 1 establishes the baseline corrective effect of LLM-mediated dialogue. A key design challenge in measuring this effect is separating a genuine misconception correction from more general belief movements that might occur for any strongly held statement following any kind of structured engagement. To address this, participants interacted with the LLM about both their most strongly held financial misconception and their most strongly held neutral statement, a belief with no objectively correct answer. If an LLM is specifically correcting financial misconceptions rather than simply inducing a general belief softening, misconceptions should shift substantially more than neutral statements. We regress the paired difference between each participant's misconception shift and neutral statement shift on an intercept to isolate the genuine correction from any diffuse response to conversational engagement.

\begin{figure}[!h]
\centering
\begin{subfigure}[t]{0.48\linewidth}
    \centering
    \includegraphics[width=\linewidth]{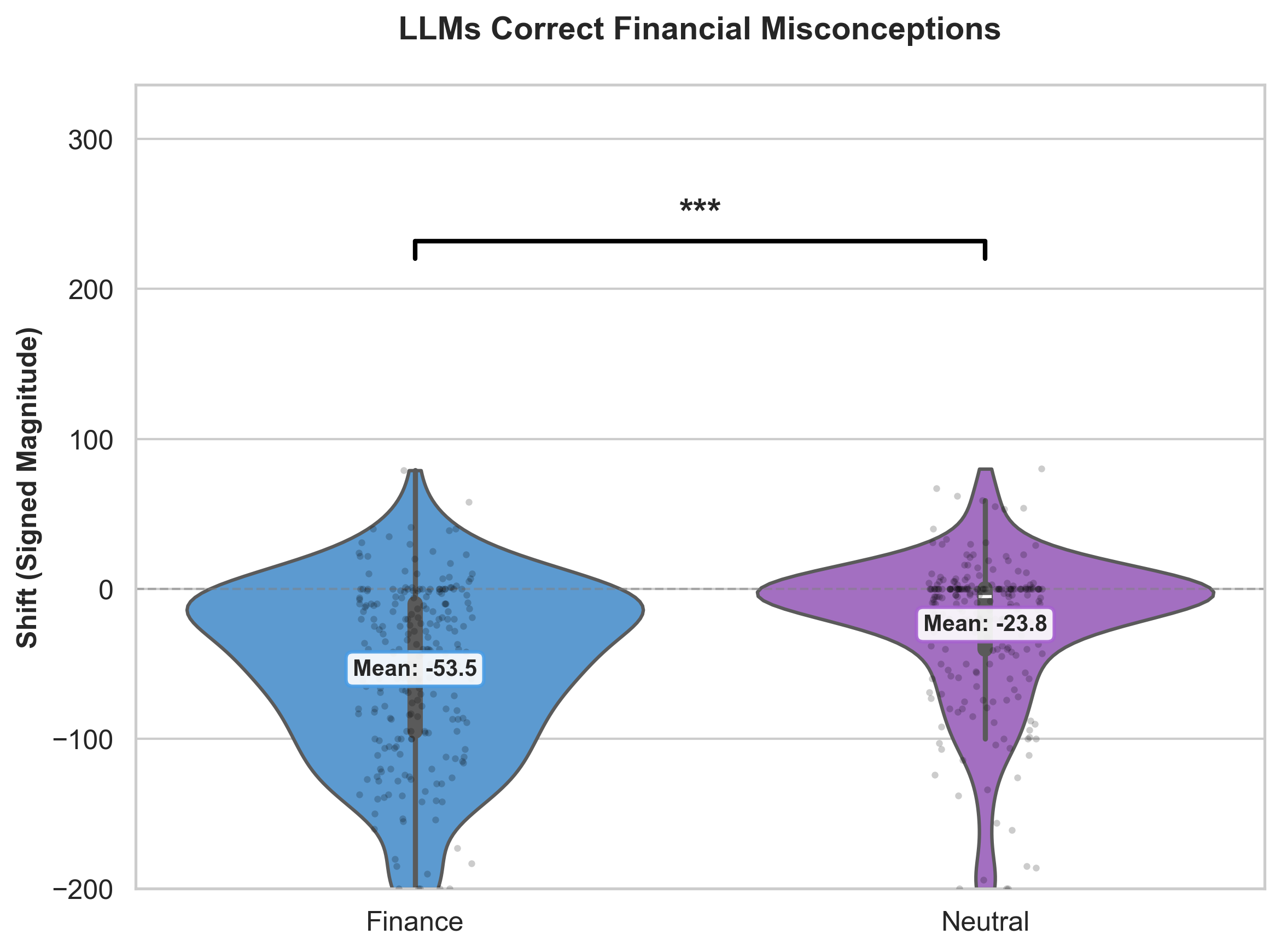}
    \caption{Immediate effect of LLM treatment.}
    \label{fig:abs_change}
\end{subfigure}%
\hfill
\begin{subfigure}[t]{0.48\linewidth}
    \centering
    \includegraphics[width=\linewidth]{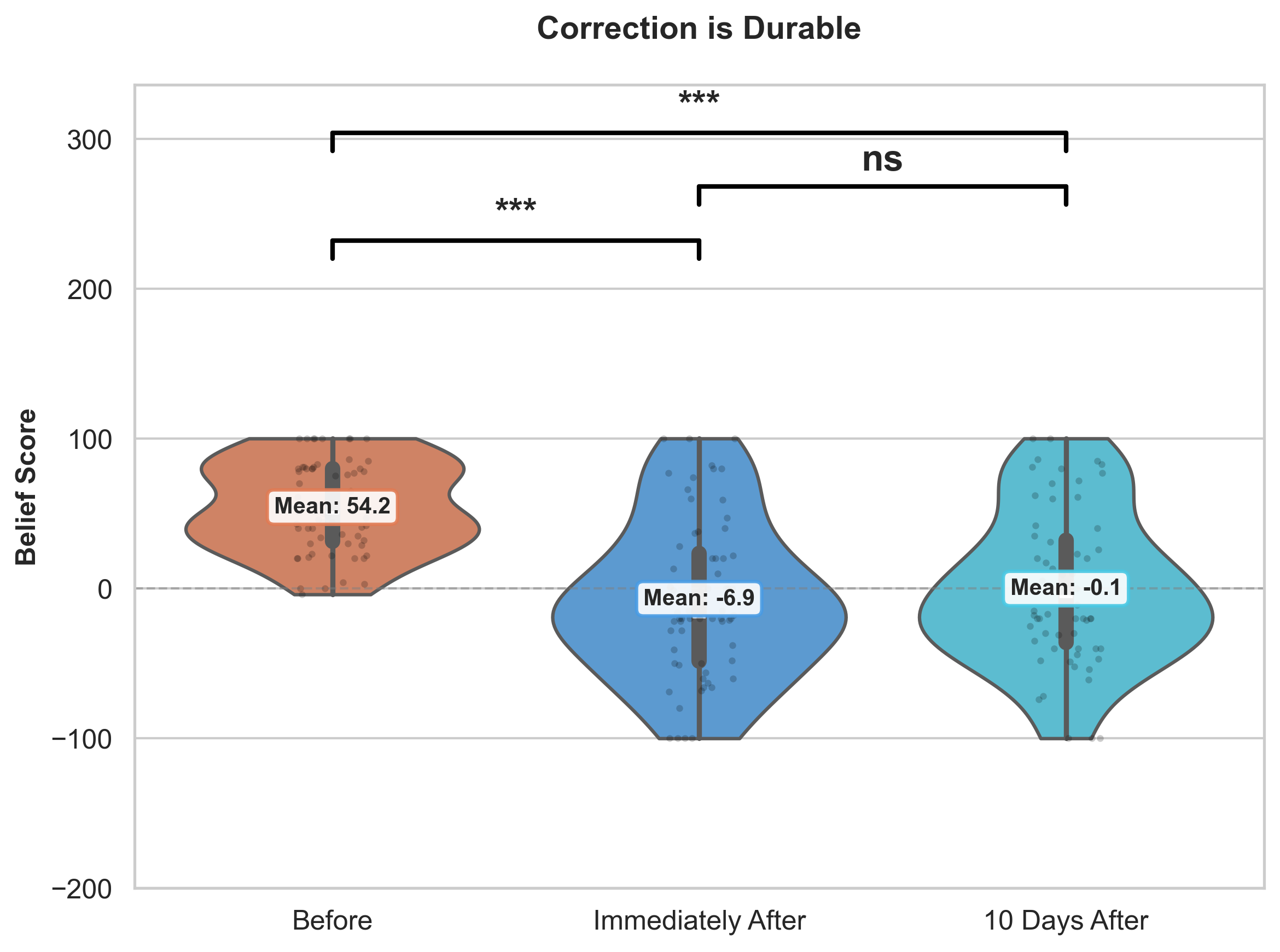}
    \caption{Longitudinal effect of LLM treatment.}
    \label{fig:longitudinal}
\end{subfigure}
\caption{Misconception correction effects. (a) Most deeply held financial misconceptions show larger absolute belief change than most deeply held neutral statements. (b) Corrections to most deeply held financial misconceptions persist over time. $^{\dagger}p<0.10$, * $p<0.05$, ** $p<0.01$, *** $p<0.001$ via paired (when comparing across conditions or time) or one-sample (when comparing belief changes or scores to 0) t-test.}
\label{fig:belief_change}
\end{figure}

As shown in Figure \ref{fig:abs_change}, participant beliefs in their most deeply held financial misconception declined by about $30$ points ($p < 0.001$; Column 2; Appendix Table \ref{tab:study1_main}) more on a scale of $-100$ to $100$ than their beliefs in neutral statements, which indicates that the chatbot was especially effective at reducing endorsement of misconceptions relative to neutral beliefs. Interestingly, the belief scores converged toward zero rather than becoming negative, which suggests that misconceptions were corrected toward neutrality rather than reversed into disagreement. We find that belief change persists 10 days after interacting with the chatbot ($p < 0.001$), as shown in Figure \ref{fig:longitudinal}. The re-recruited participant pool did not differ from the original participant pool by age, gender, education, income, or financial literacy (all $p > 0.20$; Appendix Table \ref{tab:attrition}).

We use an analysis of covariance (ANCOVA) model with clustered standard errors at the participant level to account for the mechanical relationship between baseline conviction and the room to shift. Net of this baseline effect, participant beliefs in their most deeply held  financial misconceptions still substantially declined over the neutral statements ($-17.69$, $p < 0.05$; Column 1; Appendix Table \ref{tab:study1_main}), which confirms that the corrective effect is not an artifact of regression to the mean. The interaction between pre-intervention belief strength and condition is negligible ($0.12$, $p = 0.452$; Column 1; Appendix Table \ref{tab:study1_main}), indicating that the LLM is comparably effective across the range of baseline conviction levels. 

Not all participants respond equally to the LLM intervention, however, and the pattern of heterogeneity is informative. Participants with below-average financial sophistication shifted their beliefs significantly more than their above-average counterparts, controlling for baseline conviction ($22.92$, $p < 0.05$; Column 2; Appendix Table \ref{tab:sophistication_cross}). They also generally rated the finance chatbot as more credible ($p < 0.05$, Appendix Figure \ref{fig:sophistication_credibility}), which suggests that the corrective effect of the LLM is not purely a function of argument quality but is mediated by the user's assessment of the source. A bootstrapped mediation analysis ($n_{\text{boot}} = 5{,}000$) confirms this: perceived credibility --- how favorably participants rated the finance chatbot relative to the neutral chatbot on agreeableness, convincingness, and competence -- significantly predicts belief reduction controlling for baseline conviction ($12.15$, $p < 0.001$; Column 1; Appendix Table \ref{tab:mediation}). This suggests that the corrective effect of the LLM is not purely a function of argument quality but is mediated by the user's assessment of the source. A significant indirect effect confirms that credibility partially accounts for this sophistication gap ($-2.01$, $p < 0.01$; Column 1; Appendix Table \ref{tab:mediation}). A remaining direct effect of sophistication ($-4.31$, $p < 0.05$; Column 1; Appendix Table \ref{tab:mediation}) suggests credibility is one mechanism but not the only one. This sophistication gradient raises a design question with direct practical implications: if a fixed persuasive signal is received differently depending on how credible the user finds the source, can explicitly calibrating the LLM's response complexity to the user's sophistication level close this gap? Study 3 addresses this directly.

\subsection{Correcting misconceptions requires explicit corrective intent and its absence induces entrenchment}
\label{ssec:result2}

Having established that explicitly persuasive LLM dialogue corrects financial misconceptions, we now ask what is doing the corrective work: is it the LLM's persuasive framing, the conversational dynamic it creates, or something more general about directing attention toward a held misconception? Study 2 directly tests whether corrective intent is necessary for misconception correction. We examine three conditions. In the \textit{LLM Shift} condition, the chatbot is explicitly prompted to correct the participant's misconception. In the \textit{LLM Evaluate} condition, the chatbot engages with the participant's misconception without corrective intent, which tests whether conversational interactivity alone is sufficient. In the \textit{Self-Articulate} condition, participants elaborate their own reasons for holding the misconception without any LLM interaction, which tests whether directed attention and self-reflection alone produce correction. Finally, a \textit{Distractor} condition, in which participants complete an unrelated task, serves as the reference baseline against which all three conditions are compared. This baseline measures what happens when a participant pauses and re-evaluates their prior statements with no additional intervention.

Participants whose chatbot was explicitly instructed to correct their misconceptions exhibited a belief shift of approximately 49 points relative to the \textit{Distractor} baseline ($-49.06$, $p<0.001$; Column 3; Appendix Table \ref{tab:mechanism}), robust to pre-belief controls, interaction terms, and question fixed effects. By contrast, participants who discussed their misconception with an LLM without corrective intent showed belief changes statistically indistinguishable from the \textit{Distractor} baseline (2.41, $p=0.48$; Column 3; Appendix Table \ref{tab:mechanism}), as did participants in the \textit{Self-Articulate} condition, who articulated their belief in the misconception without any LLM interaction (0.75, $p=0.816$; Column 3; Appendix Table \ref{tab:mechanism}). The equivalence of these two conditions indicates that conversational interactivity alone adds no corrective value beyond articulating a belief.

\begin{figure}[!h]
    \centering
    \includegraphics[width=0.8\linewidth]{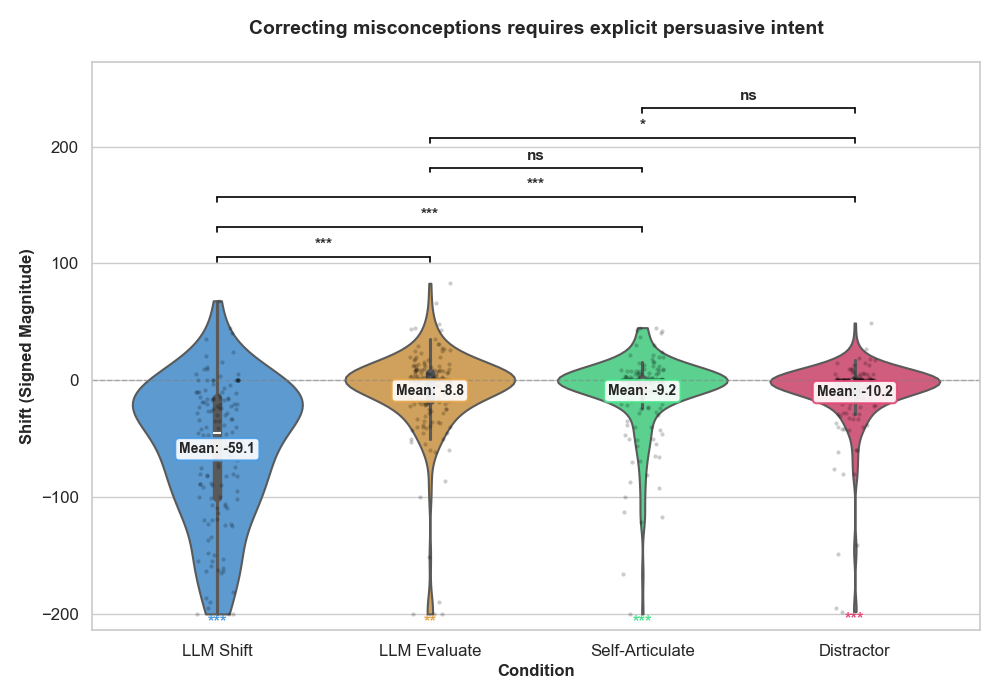}
    \caption{Violin plots of the LLM treatment when the LLM is explicitly prompted to persuade the participant (\textit{LLM Shift}), when the LLM is prompted to discuss with the participant (\textit{LLM Evaluate}), when the participant articulates their belief without any LLM interaction (\textit{Self-Articulate}), and when the participant completes a distractor task (\textit{Distractor}). $^{\dagger}p<0.10$, * $p<0.05$, ** $p<0.01$, *** $p<0.001$}
    \label{fig:mechanism}
\end{figure}

We also find that LLMs with explicit persuasive intent are especially effective for participants with stronger prior beliefs in a misconception. The corrective effect of the LLM increases with the strength of the initial misconception ($-0.60$, $p < 0.001$; Column 2; Appendix Table \ref{tab:mechanism}). No such moderation is observed for the \textit{LLM Evaluate} or \textit{Self-Articulate} conditions (both $p > 0.30$). Taken together, these results suggest that the corrective power of LLM chatbots stems specifically from their persuasive framing rather than from the conversational medium itself, and that this framing is most potent precisely when misconceptions are most deeply held.

If not designed correctly, however, LLM interaction can entrench rather than correct misconceptions. We define entrenchment as a binary outcome, equal to one if a participant moved at least 5 points further in the direction of their prior belief following the intervention, and equal to zero if not.\footnote{We find the conclusions were robust to the choice of entrenchment threshold (Appendix Table \ref{tab:entrench_sensitivity}).} We estimate the probability of entrenchment using a linear probability model (LPM) with the distractor condition as the reference group, with results confirmed by logit. 

We find that simply articulating reasons in support of a held misconception significantly increased entrenchment relative to doing nothing (8.8\%, $p < 0.05$; Column 1; Appendix Table~\ref{tab:entrench}). Interacting with an LLM regarding the misconception approximately doubled this effect (16.8\%, $p < 0.001$; Column 1; Appendix Table~\ref{tab:entrench}), raising the entrenchment rate from a baseline of 10.8\% to 27.6\%. This suggests that conversational engagement with an LLM amplifies the entrenchment that self-articulation alone already produces. By contrast, entrenchment in the \textit{LLM Shift} condition was statistically indistinguishable from the \textit{Distractor} baseline (-$2.4\%, p = 0.440$; Column 3; Appendix Table~\ref{tab:entrench}), indicating that an explicit corrective intent was sufficient to neutralize the entrenchment that non-corrective engagement and self-articulation both produced. Notably, the entrenchment effect in the \textit{LLM Evaluate} condition was significantly larger than in the \textit{Self-Articulate} condition (9.1\%, $p=0.038$). 

\begin{figure}
    \centering
    \includegraphics[width=0.9\linewidth]{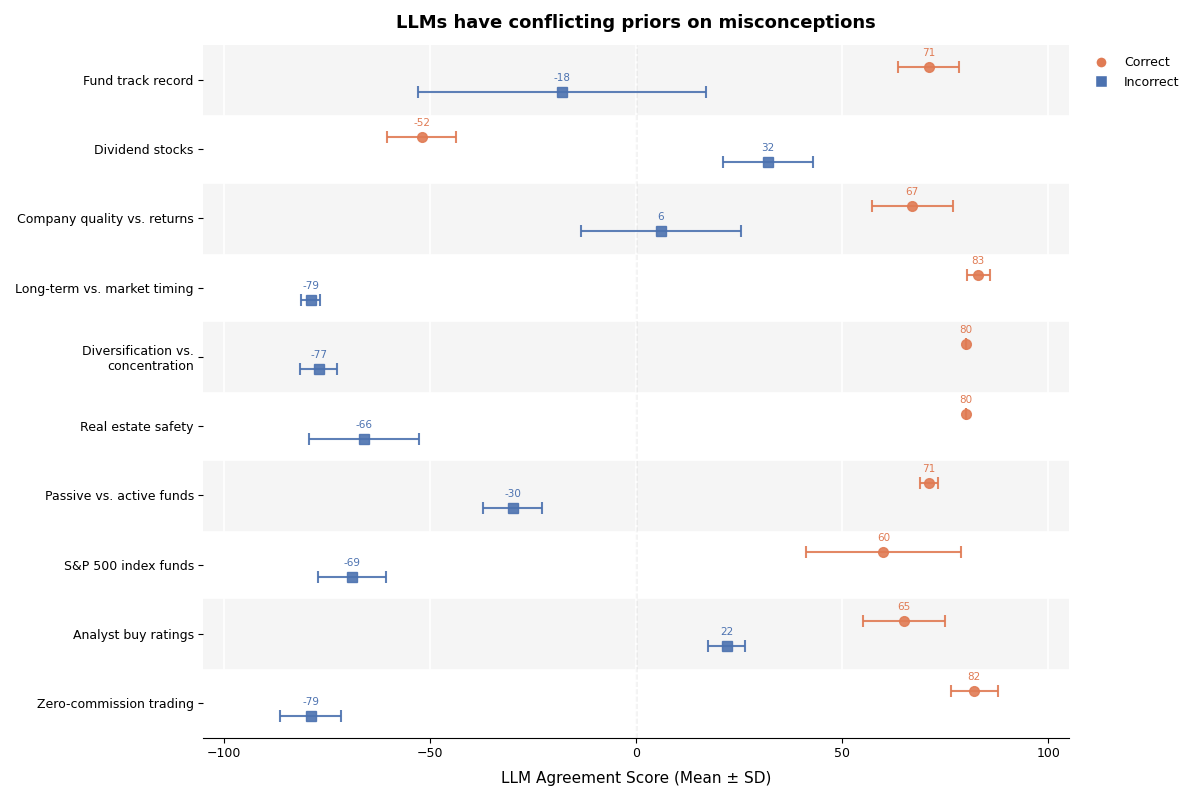}
    \caption{Language model agreement scores for 10 pairs of finance statements, where each pair contains one correct statement and one common misconception. For each statement, GPT 4o was queried 5 times and asked to rate agreement on a scale from $-100$ (strongly disagree) to $+100$ (strongly agree).}
    \label{fig:priors}
\end{figure}

These results raise a natural question: why doesn't the LLM in the \textit{LLM Evaluate} condition push back on misconceptions on its own? If the model reliably identifies these misconceptions, one would expect even undirected engagement to trend toward correction. As shown in Figure \ref{fig:priors}, entrenchment is consistent with the LLM's own inconsistent stance on these statements. The model's stance on these financial statements is noisy, with several misconceptions eliciting responses spanning both agreement and disagreement. A consistent model would produce scores symmetric around zero for each pair, strong agreement with the correct statement mirrored by strong disagreement with the misconception. Instead, we see several misconceptions receive positive agreement scores on average, the error bars showing that even within a single topic the model's stance can swing from agreement to disagreement across calls. A participant in the \textit{LLM Evaluate} condition could therefore encounter a model that actively validates the misconception it was meant to discuss. Without explicit corrective framing, the model does not reliably treat a misconception as wrong, which makes a legitimization effect possible: the act of discussing a misconception with an LLM interlocutor that neither challenges nor dismisses it may implicitly signal that the belief is reasonable, strengthening it beyond what directed self-reflection alone produces.

\subsection{More sophisticated LLM responses drive greater correction}
\label{ssec:result4}

Study 3 tests the role of recipient receptivity, whether the corrective effect of an explicitly persuasive LLM depends on the legibility and perceived credibility of its arguments. Studies 1 and 2 establish that intent is necessary for correction, but they also reveal heterogeneity. Participants with lower self-reported financial sophistication shift their beliefs more than those with higher sophistication, suggesting that the same corrective signal lands differently depending on the recipient. This raises a design question with direct practical implications: if a fixed persuasive signal is less effective for more sophisticated users, can tailoring the LLM's response complexity to the participant's sophistication level close this gap? Study 3 addresses this by comparing three conditions: an LLM whose response complexity is matched to the participant's self-reported sophistication level (\textit{Match}), one pitched above it (\textit{Mismatch High}), and one pitched below it (\textit{Mismatch Low}), with a minimum gap of three sophistication levels required in the mismatch conditions to ensure treatment intensity.

\begin{figure}[!h]
    \centering
    \includegraphics[width=0.8\linewidth]{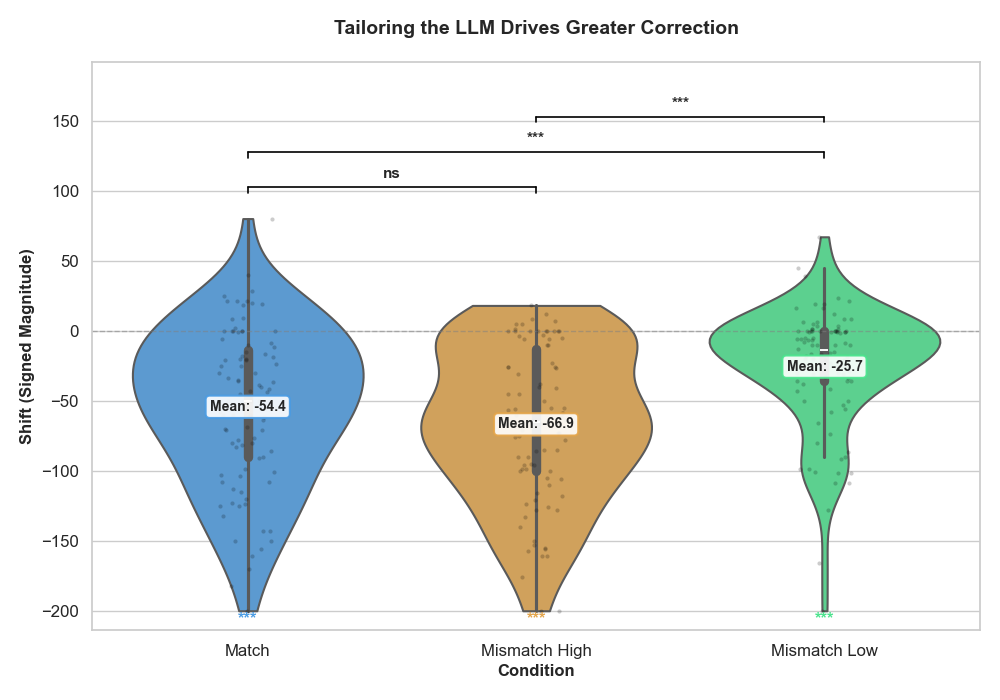}
    \caption{Violin plots of the efficacy of treatment when the LLM matches the participant's level of sophistication (\textit{Match}), is above the participant's level of sophistication (\textit{Mismatch High}), and is below the participant's level of sophistication (\textit{Mismatch Low}). $^{\dagger}p<0.10$, * $p<0.05$, ** $p<0.01$, *** $p<0.001$}
    \label{fig:sophistication_match}
\end{figure}

Sophistication matching significantly predicted the level of misconception correction. Participants who received matched arguments corrected their misconceptions by $54.39$ points on average ($p < 0.001$; Column 1; Appendix Table \ref{tab:sophistication}). Those who received above-level arguments corrected by a statistically indistinguishable amount ($-12.53$ relative to \textit{Match}, $p = 0.127$; Column 1; Appendix Table \ref{tab:sophistication}), with the point estimate trending toward greater rather than lesser correction. Those who received below-level arguments corrected substantially less, shifting $28.65$ points fewer toward correction than the \textit{Match} condition ($p < 0.001$; Column 1; Appendix Table \ref{tab:sophistication}). The effect of above-level arguments was not statistically distinguishable from the matched condition, suggesting that more complex arguments do not meaningfully improve belief updating beyond what matched arguments already achieve. These effects were robust to controlling for baseline beliefs. The sophistication mismatch effect was largely homogeneous across baseline belief levels: the interaction between \textit{Mismatch High} and prior belief was negligible ($0.06$, $p = .84$; Column 3; Appendix Table \ref{tab:sophistication}), though the interaction between \textit{Mismatch Low} and prior belief approached marginal significance ($0.49$, $p = .072$; Column 3; Appendix Table \ref{tab:sophistication}), suggesting that below-level arguments may be somewhat less effective for participants with stronger prior beliefs.

Perception of the chatbot seems to play a role. Participants who received below-level arguments consistently rated them more negatively relative to both matched and above-level participants. They found the arguments less agreeable ($p < 0.001$ vs.\ matched; $p < 0.01$ vs.\ above-level), less convincing both as a recommendation ($p < 0.001$ vs.\ both) and as an AI ($p < 0.001$ vs.\ both), and perceived the AI itself as less competent ($p < 0.001$ vs.\ both) and less useful ($p < 0.001$ vs.\ both). They also felt less encouraged by the interaction ($p < 0.01$ vs.\ matched; $p < 0.05$ vs.\ above-level). By contrast, participants who received above-level arguments were statistically indistinguishable from those who received matched arguments across nearly every dimension: agreement ($p = 0.276$), convincingness as a recommendation ($p = 0.703$), convincingness as an AI ($p = 0.582$), competence ($p = 0.851$), and usefulness ($p = 0.372$) all showed no significant difference. The one dimension that set matched participants apart from both other groups was felt challenge: those who received matched arguments reported feeling more intellectually challenged than those who received either above- or below-level arguments (both $p < 0.001$). Together, these results suggest that below-level arguments are dismissed as less credible and less useful, likely accounting for the weaker belief updating observed in that condition, while the null effect of above-level arguments relative to matched arguments may reflect competing forces (e.g., higher complexity may impede comprehension while still signaling credibility), yielding no net gain in belief updating.

\section{Discussion}
\label{sec:discussion}

Across three pre-registered experiments, we find that LLM-mediated conversations can produce large, durable corrections to financial misconceptions, but that this corrective potential depends critically on the design and deployment of the intervention. These findings speak to a broader question in behavioral economics: why financial misconceptions persist despite the ready availability of expert knowledge, and what kind of intervention can actually close the gap. Our results suggest the answer lies not in information delivery per se, but in the combination of deliberate persuasive intent, perceived source credibility, and argument legibility, conditions that conversational LLMs can satisfy at scale, but only when purposefully designed to do so.

Study 1 (Section \ref{ssec:result1}) establishes the baseline case: brief, tailored LLM conversations produce belief corrections of approximately 30 points on a $-100$ to $100$ scale, with effects persisting at least 10 days post-intervention. The convergence of post-intervention beliefs toward zero rather than into disagreement suggests that the LLM is correcting misconceptions rather than simply overwhelming participants with counterarguments, a distinction with practical implications for deployment. 

Study 2 (Section \ref{ssec:result2}) reveals the mechanism underlying this correction and, in doing so, substantially qualifies the optimism of Study 1. The core finding that explicit persuasive intent is a necessary condition for belief change, while undirected LLM engagement actively entrenches misconceptions, has a clear implication for the behavioral economics literature: it is not attention to a misconception that drives correction, but the specific form taken by that attention. This aligns with the elaboration likelihood model \citep{petty1986message}, which predicts that directing cognitive resources toward a held belief strengthens it, unless those resources are channeled into evaluating counterarguments. Crucially, entrenchment in the \textit{LLM Evaluate} condition significantly exceeded it in \textit{Self-Articulate}, ruling out elaboration volume as the sole driver of entrenchment, and pointing toward a legitimization effect: an interlocutor that engages a belief without challenging it may implicitly validate it. We cannot adjudicate this mechanism directly, but it generates a testable prediction: the entrenchment gap between \textit{LLM Evaluate} and \textit{Self-Articulate} should be mediated by perceived belief legitimacy. The practical implication is pointed regardless: financial chatbots deployed for general engagement or ``financial wellness'' without corrective intent are not neutral tools. They may systematically worsen the misconceptions they were designed to address.

Study 3 (Section \ref{ssec:result4}) extends these findings by demonstrating that the \textit{calibration} of the LLM's response complexity to the user's sophistication level materially affects correction. Below-level arguments are dismissed as less credible and produce substantially weaker belief updating, while above-level arguments are statistically indistinguishable from matched arguments in their corrective effect. This asymmetry is consistent with a credibility-signaling account: users who perceive the LLM's arguments as unsophisticated relative to their own knowledge update their assessment of the LLM's competence downward, and this credibility loss attenuates persuasion. The null effect of above-level arguments, by contrast, may reflect competing forces---e.g., higher complexity may impede comprehension while simultaneously signaling expertise---yielding no net gain over matched arguments. Together, Studies 2 and 3 establish that the corrective power of LLM-mediated dialogue is neither automatic nor uniform: it requires both explicit persuasive intent and sufficient alignment between response complexity and user sophistication.

Taken together, these findings reframe the question of what makes financial literacy interventions effective. The limiting factor is not access to accurate information, which has  been abundant, but rather the delivery conditions under which that information produces genuine belief revision. Conversational LLMs, when designed with explicit corrective intent and calibrated to the user, satisfy those conditions at a scale and cost that no prior intervention has achieved. The risk is that the same technology, deployed without those design constraints, may do the opposite.

\section{Methods}
\label{sec:methods}

The study protocol and primary analyses were preregistered prior to data collection.\footnote{asPredicted Study 241531, 268220, and 268222} All procedures were approved by the authors' institutional review board, and participants provided informed consent. 

\subsection{Chatbot Design}

We implement the chatbot intervention using a version of GPT-4o \cite{hurst2024gpt} prompted to serve as a financial expert. To achieve this goal, we construct a chatbot capable of addressing common misconceptions in personal finance with the clarity, authority, and trustworthiness of an expert advisor that remains accessible to non-specialist users. 

To ground the chatbot’s responses in genuine expert reasoning, we collect rationales from finance professors and integrate these rationales into the LLM’s prompt. This seeding process provides the model with structured, domain-relevant expert knowledge that shapes its conversational output. The design ensures that participants interact with an LLM finance expert that not only counters misconceptions, but also demonstrates the reasoning process behind expert financial advice.

These rationales comprise 10 of the most common financial misconceptions cited in the literature. The misconceptions most relevant to our study fall into three broad categories. The first concerns performance prediction illusions, or the belief that financial markets can be reliably predicted using past performance or various forecasting methods \citep{bailey2011behavioral,frazzini2008dumb,carhart1997persistence,fama2010luck,kothari2016analysts}. A second category involves risk assessment and diversification errors, which reflect misunderstandings about investment risk and the benefits of diversification \citep{benartzi2001naive,goetzmann2008equity,kumar2009hard,hartzmark2019dividend}. Finally, product structure and cost misconceptions highlight gaps in understanding how financial products function in practice, including hidden costs and structural differences that materially affect returns \citep{choi2010investors,barber2005out,elton2004spiders,gruber1996another}. For each participant, we seed the LLM’s prompt with reasoning corresponding to their most strongly held misconception.

\subsection{Statistical Methods}

Our primary dependent variable is belief shift. A naive absolute measure of belief change is insufficient for our 
purposes because it conflates correction with entrenchment: a participant who moves from $+60$ to $+40$ and one who moves from $+60$ to $+80$ produce identical absolute shifts of 20 points but represent opposite outcomes. The signed magnitude measure resolves this by encoding direction relative to the participant's prior, such that correction is always positive and reinforcement always negative. Because pre-intervention beliefs are positive by construction in our design, the measure reduces to the raw pre-post difference in practice; the signed formulation is retained for generalizability.

\begin{equation}
    \Delta_i = \text{Pre}_i - \text{Post}_i
\end{equation}
\vspace{-1em}
\begin{equation}
\tilde{\Delta}_i = \text{sign}(\text{Pre}_i) \times \Delta_i.
\end{equation}

\vspace{0.5em}

\noindent As secondary dependent variables, we measure how participants perceived the chatbot itself. After the interaction, participants rated the extent to which they agreed with the chatbot’s recommendation, as well as how convincing, likable, and competent they found it, on a Likert scale of $1$ to $7$.

\subsection{Procedure}

Across all three studies, participants were asked to rate their agreement with 10 financial misconceptions on a scale from $-100$ to $100$, without being told that these were misconceptions. They were then shown their highest-rated statement alongside the numerical rating they provided and asked to explain why they found it compelling. When applicable, this explanation initiated the LLM treatment condition. After completing the treatment, participants re-rated their belief in the same statement on the same $-100$ to $100$ scale, and evaluated the chatbot by rating their agreement with its recommendation and how convincing, likable, and competent they found it on a Likert scale of $1$ to $7$.

\noindent \textbf{Study 1.} Participants completed the same rating and interaction procedure for a set of neutral statements in addition to the financial misconceptions. To ensure that observed effects were not attributable to presentation order, participants were randomly assigned with equal probability to one of two sequences: (1) neutral statements first, followed by financial misconceptions, or (2) financial misconceptions first, followed by neutral statements.

\noindent \textbf{Study 2.} Participants were randomly assigned with equal probability to one of four conditions: (1) an LLM explicitly prompted to shift their belief (\textit{LLM Shift}); (2) an LLM prompted to engage with their belief without corrective intent (\textit{LLM Evaluate}); (3) a text box in which they articulated their own reasons for holding the misconception (\textit{Self-Articulate}); or (4) a distractor task requiring them to define various financial terms (\textit{Distractor}).

\noindent \textbf{Study 3.} Participants were randomly assigned to interact with an LLM calibrated to one of three sophistication levels relative to their own self-reported financial sophistication: (1) matched to their level (\textit{Match}); (2) pitched above their level (\textit{Mismatch High}); or (3) pitched below their level (\textit{Mismatch Low}). To ensure the treatment was sufficiently strong, assignment to a mismatch condition required a minimum gap of three sophistication levels between the participant and the LLM.

\subsection{Participants}

For all three studies, participants on Prolific were recruited to complete the task online on Qualtrics. Participation across all studies was limited to adults aged 25--65 who were fluent in English, resided in the United States, and had full-time employment status. Each participant was paid at a rate of \$12 per hour and could only complete any of the three studies once.

\noindent \textbf{Study 1.} A total of 423 participants were recruited. After applying pre-registered exclusion criteria, including attention checks, minimum completion time ($>$180s), outlier removal, and a relevance filter requiring at least 3 on-topic conversational messages, 239 participants remained (57\% retention; Appendix Table~\ref{tab:sample_study1}).

\noindent \textbf{Study 2.} A total of 785 participants were recruited across four conditions (\textit{LLM Shift}: $n=189$; \textit{LLM Evaluate}: $n=200$; \textit{Self-Articulate}: $n=198$; \textit{Distractor}: $n=198$). After applying pre-registered exclusion criteria, 721 participants remained across conditions (\textit{LLM Shift}: $n=144$, 76\%; \textit{LLM Evaluate}: $n=198$, 99\%; \textit{Self-Articulate}: $n=191$, 96\%; \textit{Distractor}: $n=188$, 95\%; Appendix Table~\ref{tab:sample_study2}). The relevance filter was applied only to the Shift condition, as only that condition involved multi-turn LLM conversation.

\noindent \textbf{Study 3.} A total of 477 participants were recruited across three conditions (Match: $n=222$; Mismatch High: $n=130$; Mismatch Low: $n=125$). After applying pre-registered exclusion criteria, 289 participants remained (Match: $n=97$, 44\%; Mismatch High: $n=92$, 71\%; Mismatch Low: $n=100$, 80\%; Appendix Table~\ref{tab:sample_study3}).

\section{Limitations}

Several limitations qualify our findings. First, its external validity is limited: the study relies on a U.S.-based Prolific sample and a single LLM, so effects may not generalize across populations, market conditions, or model updates. Second, outcomes are based on self-reported belief ratings rather than observed financial behavior, which raises concerns about demand effects, regression to the mean, and the durability of belief change. Third, our design selected each participant’s strongest statement, which improves the statistical power of our findings, but may overstate the average effects relative to more typical beliefs. Fourth, the minimum chat length requirement could mask heterogeneity in intervention intensity. Fifth, while our follow-up suggests persistence in effects, it is based on a subset of participants and may reflect selective attrition. 

Future work should address these limitations by testing chatbot interventions with more diverse samples and contexts, extending the follow-up periods to assess the long-term durability of belief change, and linking belief shifts to actual financial behaviors such as portfolio allocation or savings decisions. Comparative studies across different LLM architectures and prompting strategies could also clarify the role of model design in shaping outcomes. Moreover, hybrid interventions in which chatbots are paired with human financial educators may combine the scale of AI with the accountability of expert oversight. Understanding how to calibrate chatbot persuasiveness and tailor interventions to user needs without overstepping into prescriptive financial advice remains an open and important research direction.

\bibliography{sn-bibliography}

\newpage

\appendix
\section{Appendix}

\subsection{Study 1 Results and Analysis}

\begin{equation}
\text{BeliefShift}_i = \beta_0 + \beta_1 \mathbf{1}[\text{Shift}]_i + \beta_2 \mathbf{1}[\text{Evaluate}]_i + \beta_3 \mathbf{1}[\text{Self-Articulate}]_i + \varepsilon_i
\end{equation}

\begin{equation}
\text{BeliefShift}_i = \beta_0 + \beta_1 \mathbf{1}[\text{Shift}]_i + \beta_2 \mathbf{1}[\text{Evaluate}]_i + \beta_3 \mathbf{1}[\text{Self-Articulate}]_i + \beta_4 \text{Initial}_i + \varepsilon_i
\end{equation}

\begin{equation}
\begin{aligned}
\text{BeliefShift}_i = \, & \beta_0 + \beta_1 \mathbf{1}[\text{Shift}]_i + \beta_2 \mathbf{1}[\text{Evaluate}]_i + \beta_3 \mathbf{1}[\text{Self-Articulate}]_i + \beta_4 \text{Initial}_i \\
& + \beta_5 \left(\mathbf{1}[\text{Shift}]_i \times \text{Initial}_i\right) \\
& + \beta_6 \left(\mathbf{1}[\text{Evaluate}]_i \times \text{Initial}_i\right) \\
& + \beta_7 \left(\mathbf{1}[\text{Self-Articulate}]_i \times \text{Initial}_i\right) + \varepsilon_i
\end{aligned}
\end{equation}

\begin{table}[!h]
\centering
\caption{OLS: Misconception Correction}
\begin{tabular}{lcccc}
\hline
 & \multicolumn{2}{c}{(1)} & \multicolumn{2}{c}{(2)} \\
 & Coef. & SE & Coef. & SE \\
\hline
Intercept                          & $-2.40$       & $6.84$ &                &        \\
Finance                            & $-17.69^{**}$ & $8.88$ &                &        \\
Initial                                & $0.50^{***}$  & $0.11$ &                &        \\
Finance $\times$ Initial               & $0.12$        & $0.16$ &                &        \\
\hline
Paired Difference ($\Delta_{\text{fin}} - \Delta_{\text{neu}}$) & & & $-29.72^{***}$ & $4.45$ \\
\hline
$N$                                & \multicolumn{2}{c}{$482$}   & \multicolumn{2}{c}{$241$}   \\
$R^2$                              & \multicolumn{2}{c}{$0.110$} & \multicolumn{2}{c}{$0.000$} \\
Adj.\ $R^2$                        & \multicolumn{2}{c}{$0.104$} & \multicolumn{2}{c}{$0.000$} \\
\hline
\multicolumn{5}{l}{$^{***}\ p < 0.001$, $^{**}\ p < 0.05$, $^{*}\ p < 0.10$.} \\
\multicolumn{5}{l}{Column (1): cluster-robust standard errors. Column (2): conventional standard errors.} \\
\multicolumn{5}{l}{Column (1) dependent variable: post-intervention belief rating. Reference group: neutral statements.} \\
\multicolumn{5}{l}{Column (2) dependent variable: paired difference in belief shift (finance minus neutral).} \\
\end{tabular}
\label{tab:study1_main}
\end{table}

\noindent \textbf{Perception Analysis.} To test whether perceived credibility mediates the relationship between financial sophistication and belief correction, we estimate a bootstrapped mediation model ($n_{\text{boot}} = 5{,}000$) with self-reported financial sophistication (1--7 scale) as the independent variable ($X$), perceived credibility advantage as the mediator ($M$), and finance belief shift as the outcome ($Y$), controlling for baseline conviction throughout. Perceived credibility advantage is a composite measure computed as the difference between participants' ratings of the finance and neutral chatbots on agreeableness, convincingness, and competence, where a positive value indicates the finance chatbot was rated more favorably. Finance belief shift is defined as the pre-intervention minus post-intervention belief score, such that positive values indicate corrective movement away from the misconception. The indirect effect (the portion of the sophistication $\to$ shift relationship that operates through credibility) is estimated as the product of the $X \to M$ and $M \to Y$ path coefficients, with significance assessed via the bootstrapped confidence interval. A significant indirect effect whose confidence interval excludes zero indicates that credibility is a meaningful pathway; a remaining direct effect indicates that sophistication predicts correction through additional mechanisms beyond credibility.

\begin{figure}
    \centering
    \includegraphics[width=\linewidth]{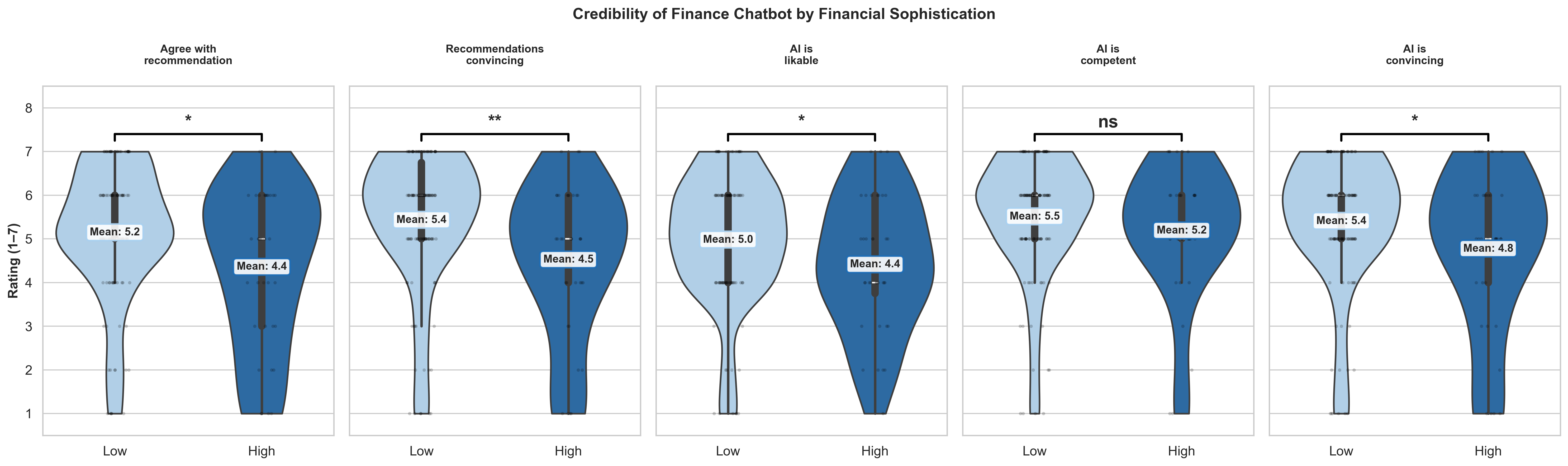}
    \caption{Participants with low financial sophistication agree more with the chatbot's recommendation, find the chatbot and its recommendation more convincing, and find the chatbot more likable. $^{\dagger}p<0.10$, * $p<0.05$, ** $p<0.01$, *** $p<0.001$}
    \label{fig:sophistication_credibility}
\end{figure}

\begin{table}[!h]
\centering
\caption{OLS: Sophistication Cross-Section Misconception Correction}
\label{tab:sophistication_cross}
\begin{tabular}{lcccc}
\hline
 & \multicolumn{2}{c}{(1)} & \multicolumn{2}{c}{(2)} \\
 & Coef. & SE & Coef. & SE \\
\hline
High sophistication & $16.69$ & $10.88$ & $22.92^{*}$ & $10.43$ \\
Baseline conviction &         &         & $-0.59^{***}$ & $0.13$ \\
Intercept           & $-57.49^{***}$ & $4.87$ & $-26.96^{***}$ & $7.50$ \\
\hline
$N$        & \multicolumn{2}{c}{$162$}   & \multicolumn{2}{c}{$162$}   \\
$R^{2}$    & \multicolumn{2}{c}{$0.016$} & \multicolumn{2}{c}{$0.130$} \\
Adj.\ $R^2$ & \multicolumn{2}{c}{$0.010$} & \multicolumn{2}{c}{$0.119$} \\
\hline
\multicolumn{5}{l}{$^{***}\ p < 0.001$, $^{**}\ p < 0.01$, $^{*}\ p < 0.05$.} \\
\multicolumn{5}{l}{Dependent variable: finance belief shift (pre $-$ post; positive $=$ corrective).} \\
\multicolumn{5}{l}{Reference category: below-average sophistication. Sample excludes participants} \\
\multicolumn{5}{l}{reporting average sophistication. Heteroscedasticity-robust SEs (HC3).} \\
\end{tabular}
\end{table}

\begin{table}[!h]
\centering
\caption{Mediation Analysis: Sophistication $\to$ Perceived Credibility $\to$ Misconception Correction}
\label{tab:mediation}
\begin{tabular}{lcc}
\hline
 & \multicolumn{2}{c}{(1)} \\
 & Coef. & SE \\
\hline
\multicolumn{3}{l}{\textit{Path coefficients}} \\
$X \to M$: Sophistication $\to$ Credibility & $-0.176^{**}$ & $0.065$ \\
$M \to Y$: Credibility $\to$ Belief shift   & $12.15^{***}$ & $2.01$ \\
\hline
\multicolumn{3}{l}{\textit{Effect decomposition}} \\
Total effect   & $-6.31^{**}$  & $2.15$ \\
Direct effect  & $-4.31^{*}$   & $2.06$ \\
Indirect effect & $-2.01^{**}$ & $0.88$ \\
\hline
$N$ & \multicolumn{2}{c}{$241$} \\
\hline
\multicolumn{3}{l}{$^{***}\ p < 0.001$, $^{**}\ p < 0.01$, $^{*}\ p < 0.05$.} \\
\multicolumn{3}{l}{$X$ = self-reported financial sophistication (1--7 scale); $M$ = perceived credibility} \\
\multicolumn{3}{l}{advantage (finance minus neutral chatbot composite of agreeableness,} \\
\multicolumn{3}{l}{convincingness, and competence); $Y$ = finance belief shift (pre $-$ post;} \\
\multicolumn{3}{l}{positive $=$ corrective). Covariate: baseline finance belief. Indirect effect} \\
\multicolumn{3}{l}{SE and significance assessed via $n_{\text{boot}} = 5{,}000$ bootstrap samples.} \\
\end{tabular}
\end{table}

\newpage

\subsection{Study 2 Results and Analysis}

\begin{table}[!h]
\centering
\caption{OLS: Mechanism Behind Misconception Correction}
\begin{tabular}{lcccccc}
\hline
 & \multicolumn{2}{c}{(1)} & \multicolumn{2}{c}{(2)} & \multicolumn{2}{c}{(3)} \\
 & Coef. & SE & Coef. & SE & Coef. & SE \\
\hline
Intercept (Distractor)      & $6.01^{\dagger}$  & $3.28$ & $-2.59$        & $3.77$ & $-3.06$        & $4.75$ \\
LLM Evaluate                & $1.89$            & $3.41$ & $8.11$         & $6.58$ & $2.41$         & $3.42$ \\
LLM Shift                   & $-48.63^{***}$    & $5.19$ & $-14.35$       & $9.15$ & $-49.06^{***}$ & $5.01$ \\
Self-Articulate             & $0.06$            & $3.20$ & $3.00$         & $5.15$ & $0.75$         & $3.23$ \\
Initial                     & $-0.29^{***}$     & $0.05$ & $-0.13^{\dagger}$ & $0.08$ & $-0.27^{***}$  & $0.05$ \\
LLM Evaluate $\times$ Initial &                 &        & $-0.11$        & $0.13$ &                &        \\
LLM Shift $\times$ Initial    &                 &        & $-0.60^{***}$  & $0.18$ &                &        \\
Self-Articulate $\times$ Initial &              &        & $-0.05$        & $0.11$ &                &        \\
\hline
$N$          & \multicolumn{2}{c}{$716$}   & \multicolumn{2}{c}{$716$}   & \multicolumn{2}{c}{$716$}   \\
$R^2$        & \multicolumn{2}{c}{$0.244$} & \multicolumn{2}{c}{$0.267$} & \multicolumn{2}{c}{$0.291$} \\
Adj.\ $R^2$  & \multicolumn{2}{c}{$0.239$} & \multicolumn{2}{c}{$0.260$} & \multicolumn{2}{c}{$0.278$} \\
AIC          & \multicolumn{2}{c}{$7277$}  & \multicolumn{2}{c}{$7260$}  & \multicolumn{2}{c}{$7249$}  \\
Initial Control  & \multicolumn{2}{c}{Yes} & \multicolumn{2}{c}{Yes}     & \multicolumn{2}{c}{Yes}     \\
Interactions     & \multicolumn{2}{c}{No}  & \multicolumn{2}{c}{Yes}     & \multicolumn{2}{c}{No}      \\
Question FE      & \multicolumn{2}{c}{No}  & \multicolumn{2}{c}{No}      & \multicolumn{2}{c}{Yes}     \\
\hline
\multicolumn{7}{l}{$^{\dagger}\ p < 0.10$, $^{*}\ p < 0.05$, $^{**}\ p < 0.01$, $^{***}\ p < 0.001$.} \\
\multicolumn{7}{l}{Heteroskedasticity-robust standard errors (HC3). Distractor is the reference group.} \\
\multicolumn{7}{l}{Question FE coefficients suppressed. Reference question: Actively Managed Funds.} \\
\end{tabular}
\label{tab:mechanism}
\end{table}

\begin{equation}
\text{Entrench}_i = \beta_0 + \beta_1 \mathbf{1}[\text{Shift}]_i + \beta_2 \mathbf{1}[\text{Evaluate}]_i + \beta_3 \mathbf{1}[\text{Self-Articulate}]_i + \varepsilon_i
\end{equation}

\begin{equation}
\text{Entrench}_i = \beta_0 + \beta_1 \mathbf{1}[\text{Shift}]_i + \beta_2 \mathbf{1}[\text{Evaluate}]_i + \beta_3 \mathbf{1}[\text{Self-Articulate}]_i + \beta_4 \text{Initial}_i + \varepsilon_i
\end{equation}

\begin{equation}
\begin{aligned}
\text{Entrench}_i = \, & \beta_0 + \beta_1 \mathbf{1}[\text{Shift}]_i + \beta_2 \mathbf{1}[\text{Evaluate}]_i + \beta_3 \mathbf{1}[\text{Self-Articulate}]_i + \beta_4 \text{Initial}_i \\
& + \beta_5 \left(\mathbf{1}[\text{Shift}]_i \times \text{Initial}_i\right) \\
& + \beta_6 \left(\mathbf{1}[\text{Evaluate}]_i \times \text{Initial}_i\right) \\
& + \beta_7 \left(\mathbf{1}[\text{Self-Articulate}]_i \times \text{Initial}_i\right) + \varepsilon_i
\end{aligned}
\end{equation}

\noindent where $\text{Entrench}_i \in \{0,1\}$ equals one if participant $i$ moved at least 5 points further in the direction of their prior belief following the intervention, and zero otherwise. The distractor condition serves as the reference group. Standard errors are heteroskedasticity-robust (HC3).

\begin{table}[!h]
\centering
\caption{Linear Probability Model: Entrenchment}
\label{tab:entrench}
\begin{tabular}{lcccccc}
\hline
 & \multicolumn{2}{c}{(1)} & \multicolumn{2}{c}{(2)} & \multicolumn{2}{c}{(3)} \\
 & Coef. & SE & Coef. & SE & Coef. & SE \\
\hline
Intercept (Distractor)  & $0.108^{***}$ & $0.023$ & $0.233^{***}$ & $0.036$ & $0.180^{***}$ & $0.045$ \\
LLM Evaluate            & $0.168^{***}$ & $0.039$ & $0.172^{***}$ & $0.039$ & $0.178^{***}$ & $0.040$ \\
LLM Shift               & $-0.025$      & $0.032$ & $-0.023$      & $0.032$ & $-0.024$      & $0.032$ \\
Self-Articulate         & $0.088^{*}$   & $0.037$ & $0.081^{*}$   & $0.037$ & $0.088^{*}$   & $0.037$ \\
Initial                 &               &         & $-0.002^{***}$& $0.000$ & $-0.002^{***}$& $0.000$ \\
\hline
$N$         & \multicolumn{2}{c}{$716$}   & \multicolumn{2}{c}{$716$}   & \multicolumn{2}{c}{$716$}   \\
$R^2$       & \multicolumn{2}{c}{$0.041$} & \multicolumn{2}{c}{$0.076$} & \multicolumn{2}{c}{$0.100$} \\
Adj.\ $R^2$ & \multicolumn{2}{c}{$0.037$} & \multicolumn{2}{c}{$0.070$} & \multicolumn{2}{c}{$0.083$} \\
AIC         & \multicolumn{2}{c}{$614.1$} & \multicolumn{2}{c}{$589.4$} & \multicolumn{2}{c}{$588.2$} \\
Initial Control & \multicolumn{2}{c}{No}  & \multicolumn{2}{c}{Yes}     & \multicolumn{2}{c}{Yes}     \\
Question FE     & \multicolumn{2}{c}{No}  & \multicolumn{2}{c}{No}      & \multicolumn{2}{c}{Yes}     \\
\hline
\multicolumn{7}{l}{$^{\dagger}\ p < 0.10$, $^{*}\ p < 0.05$, $^{**}\ p < 0.01$, $^{***}\ p < 0.001$.} \\
\multicolumn{7}{l}{Heteroskedasticity-robust standard errors (HC3). Distractor is the reference group.} \\
\multicolumn{7}{l}{Question FE coefficients suppressed. Reference: Actively Managed Funds.} \\
\multicolumn{7}{l}{Logit estimates consistent in sign and significance (Appendix).} \\
\end{tabular}
\end{table}

\noindent \textbf{Entrenchment Sensitivity Analysis.} The entrenchment results in Section~\ref{ssec:result2} define entrenchment as a belief movement of at least 5 points in the direction of the prior belief. To verify that our findings are not artifacts of this threshold choice, we re-estimate the linear probability model with a pre-belief control across thresholds ranging from 1 to 20 points. As shown in Table~\ref{tab:entrench_sensitivity}, the pattern of results is stable across all thresholds: the \textit{LLM Evaluate} coefficient is positive and significant at every threshold tested, the \textit{LLM Shift} coefficient is consistently near zero and non-significant, and the \textit{Self-Articulate} coefficient is positive and significant at lower thresholds, attenuating at higher ones. The core finding that undirected LLM engagement produces the largest entrenchment effect is robust to the choice of threshold.

\begin{table}[!h]
\centering
\caption{Entrenchment Sensitivity to Threshold Definition (LPM)}
\label{tab:entrench_sensitivity}
\begin{tabular}{ccccccccc}
\hline
 & & \multicolumn{2}{c}{LLM Shift} 
   & \multicolumn{2}{c}{LLM Evaluate} 
   & \multicolumn{2}{c}{Self-Articulate} \\
Threshold & Overall Rate & Coef. & Sig. & Coef. & Sig. & Coef. & Sig. \\
\hline
1  & 0.236 & $-0.070$ &    & $+0.225$ & $***$ & $+0.102$ & $*$   \\
2  & 0.214 & $-0.048$ &    & $+0.211$ & $***$ & $+0.097$ & $*$   \\
3  & 0.198 & $-0.032$ &    & $+0.202$ & $***$ & $+0.097$ & $*$   \\
5  & 0.172 & $-0.023$ &    & $+0.172$ & $***$ & $+0.081$ & $*$   \\
7  & 0.141 & $-0.004$ &    & $+0.169$ & $***$ & $+0.077$ & $*$   \\
10 & 0.113 & $+0.006$ &    & $+0.123$ & $***$ & $+0.052$ &       \\
15 & 0.075 & $+0.012$ &    & $+0.083$ & $**$  & $+0.048$ &       \\
20 & 0.056 & $+0.032$ &    & $+0.089$ & $***$ & $+0.054$ & $**$  \\
\hline
\multicolumn{9}{l}{$^{***}\ p < 0.001$, $^{**}\ p < 0.01$, $^{*}\ p < 0.05$. 
    Distractor is the reference group.} \\
\multicolumn{9}{l}{LPM 2 includes pre-belief control (HC3 standard errors).
    Threshold indicates minimum belief} \\
\multicolumn{9}{l}{movement in the direction of the prior required to classify 
    as entrenched.} \\
\end{tabular}
\end{table}

\newpage

\subsection{Study 3 Results and Analysis}

\begin{table}[!h]
\centering
\caption{OLS: Misconception Correction by Sophistication Matching}
\label{tab:sophistication}
\begin{tabular}{lcccccc}
\hline
 & \multicolumn{2}{c}{(1)} & \multicolumn{2}{c}{(2)} & \multicolumn{2}{c}{(3)} \\
 & Coef. & SE & Coef. & SE & Coef. & SE \\
\hline
Intercept (Match)              & $-54.39^{***}$ & $5.79$ & $-31.08^{***}$ & $7.44$ & $-19.09$       & $13.19$ \\
Mismatch High                  & $-12.53$       & $8.21$ & $-11.15$       & $7.84$ & $-13.92$       & $16.18$ \\
Mismatch Low                   & $28.65^{***}$  & $7.27$ & $26.47^{***}$  & $7.12$ & $1.36$         & $15.08$ \\
Initial                        &                &        & $-0.43^{***}$  & $0.11$ & $-0.65^{**}$   & $0.23$  \\
Mismatch High $\times$ Initial &                &        &                &        & $0.06$         & $0.30$  \\
Mismatch Low $\times$ Initial  &                &        &                &        & $0.49^{\dagger}$ & $0.27$ \\
\hline
$N$         & \multicolumn{2}{c}{$283$}   & \multicolumn{2}{c}{$283$}   & \multicolumn{2}{c}{$283$}   \\
$R^2$       & \multicolumn{2}{c}{$0.102$} & \multicolumn{2}{c}{$0.157$} & \multicolumn{2}{c}{$0.172$} \\
Adj.\ $R^2$ & \multicolumn{2}{c}{$0.095$} & \multicolumn{2}{c}{$0.148$} & \multicolumn{2}{c}{$0.157$} \\
AIC         & \multicolumn{2}{c}{$3039$}  & \multicolumn{2}{c}{$3023$}  & \multicolumn{2}{c}{$3022$}  \\
Initial Control  & \multicolumn{2}{c}{No}  & \multicolumn{2}{c}{Yes} & \multicolumn{2}{c}{Yes} \\
Interactions     & \multicolumn{2}{c}{No}  & \multicolumn{2}{c}{No}  & \multicolumn{2}{c}{Yes} \\
\hline
\multicolumn{7}{l}{$^{\dagger}\ p < 0.10$, $^{*}\ p < 0.05$, $^{**}\ p < 0.01$, $^{***}\ p < 0.001$.} \\
\multicolumn{7}{l}{Heteroskedasticity-robust standard errors (HC3). Match is the reference group.} \\
\multicolumn{7}{l}{Dependent variable is belief shift (signed magnitude).} \\
\end{tabular}
\end{table}

\newpage
\newpage

\subsection{Dataset Statistics}

\textbf{Filtering.} We applied four sequential exclusion criteria to ensure data quality. First, we excluded participants who incorrectly answered the attention check question. Second, we excluded participants who completed the entire survey in under three minutes, a threshold indicating insufficient engagement with the task; this criterion did not further reduce the sample. Third, we excluded participants whose survey completion time fell more than 2.5 standard deviations above or below the mean across all conditions. Finally, where applicable, we excluded participants who did not exchange a minimum of three relevant messages with the LLM where relevance was assessed via LLM coding.

\begin{table}[!h]
\centering
\caption{Sample Construction: Study 1}
\label{tab:sample_study1}
\begin{tabular}{lc}
\hline
Step & All Participants \\
\hline
Raw                        & 423 (100\%) \\
Attention check            & 423 (100\%) \\
Min time ($>$180s)         & 423 (100\%) \\
2.5-SD outlier removal     & 417 (99\%)  \\
Relevance ($\geq$3 msgs)   & 239 (57\%)  \\
\hline
\textbf{Final $N$}         & \textbf{239 (57\%)} \\
\hline
\multicolumn{2}{l}{\small Relevance assessed by LLM rating of conversation quality.} \\
\end{tabular}
\end{table}

\begin{table}[!h]
\centering
\caption{Longitudinal Attrition: Study 1}
\label{tab:attrition}
\begin{tabular}{lcccc}
\hline
 & Re-recruited & Dropped & & \\
Variable & ($n = 114$) & ($n = 164$) & Stat & $p$ \\
\hline
Age (mean, SD)              & 42.95 (10.45) & 41.88 (10.05) & $U$ & 0.432 \\
Gender (\% female)          & 42.1 & 44.5 & $\chi^2(3)$ & 0.705 \\
Education (\% bachelor's+)  & 74.5 & 66.4 & $\chi^2(5)$ & 0.209 \\
Income (\% \$75k+)          & 51.7 & 41.5 & $\chi^2(4)$ & 0.387 \\
Financial literacy (\% above average) & 19.3 & 14.2 & $\chi^2(6)$ & 0.604 \\
\hline
\multicolumn{5}{l}{\small Continuous: Mann--Whitney $U$; categorical: $\chi^2$. Return rate: $114/278 = 41.0\%$.} \\
\multicolumn{5}{l}{\small All differences non-significant at $p > 0.20$.} \\
\end{tabular}
\end{table}

\begin{table}[!h]
\centering
\caption{Sample Construction: Study 2}
\label{tab:sample_study2}
\begin{tabular}{lcccc}
\hline
Step & Shift & Eval & Self-Articulate & None \\
\hline
Raw                      & 189 (100\%) & 200 (100\%) & 198 (100\%) & 198 (100\%) \\
Attention check          & 189 (100\%) & 200 (100\%) & 198 (100\%) & 198 (100\%) \\
Min time ($>$180s)       & 189 (100\%) & 200 (100\%) & 196 (99\%)  & 197 (99\%)  \\
2.5-SD outlier removal   & 185 (98\%)  & 198 (99\%)  & 191 (96\%)  & 188 (95\%)  \\
Relevance ($\geq$3 msgs) & 144 (76\%)  & N/A         & N/A         & N/A         \\
\hline
\textbf{Final $N$}       & \textbf{144 (76\%)} & \textbf{198 (99\%)} & \textbf{191 (96\%)} & \textbf{188 (95\%)} \\
\hline
\multicolumn{5}{l}{\small Relevance filter applied only to Shift condition (interactive LLM).} \\
\multicolumn{5}{l}{\small N/A indicates condition did not involve multi-turn conversation.} \\
\end{tabular}
\end{table}

\begin{table}[!h]
\centering
\caption{Sample Construction: Study 3}
\label{tab:sample_study3}
\begin{tabular}{lccc}
\hline
Step & Match & Mismatch High & Mismatch Low \\
\hline
Raw                      & 222 (100\%) & 130 (100\%) & 125 (100\%) \\
Attention check          & 222 (100\%) & 130 (100\%) & 125 (100\%) \\
Min time ($>$180s)       & 221 (100\%) & 130 (100\%) & 125 (100\%) \\
2.5-SD outlier removal   & 214 (96\%)  & 125 (96\%)  & 119 (95\%)  \\
Relevance ($\geq$3 msgs) & 97 (44\%)   & 92 (71\%)   & 100 (80\%)  \\
\hline
\textbf{Final $N$}       & \textbf{97 (44\%)} & \textbf{92 (71\%)} & \textbf{100 (80\%)} \\
\hline
\multicolumn{4}{l}{\small Percentages relative to raw $N$ within each condition.} \\
\end{tabular}
\end{table}

\newpage

\begin{table}[!h]
\centering
\caption{Study 1}
\label{tab:prebelief_study1}
\begin{tabular}{lrrr}
\hline
Statement & Mean & SD & $N$ \\
\hline
Walking is an overrated form of exercise.                           & 14.00 & 8.49  & 2  \\
Stocks that Wall Street analysts rate as `buy' will outperform\dots & 33.00 & 12.19 & 4  \\
For most people, concentrating your money in a few stocks\dots      & 41.80 & 18.02 & 5  \\
People are less productive when they work from home.                & 44.50 & 33.46 & 10 \\
Real estate is always a safe investment regardless of\dots          & 45.69 & 22.78 & 13 \\
Watching TV before bed makes you fall asleep slower.                & 48.33 & 25.11 & 18 \\
Cats are better pets than dogs.                                     & 50.11 & 41.88 & 9  \\
Better-run companies are always better stocks to buy\dots           & 50.14 & 26.45 & 22 \\
All S\&P 500 index funds are interchangeable.                       & 51.50 & 68.59 & 2  \\
Actively managed funds tend to earn higher returns\dots             & 53.63 & 27.23 & 19 \\
Zero-commission trades means that trading is costless.              & 53.88 & 52.64 & 8  \\
When choosing between similar funds, pick the fund\dots             & 53.94 & 34.78 & 18 \\
You can learn more from school than from YouTube.                   & 56.05 & 28.50 & 40 \\
It's better to work out in the evening than in the morning.         & 56.50 & 34.10 & 16 \\
It is better to invest in stocks that pay dividends\dots            & 58.12 & 33.39 & 17 \\
It is not useful to learn how to play a musical instrument.         & 60.00 & ---   & 1  \\
I do not like cooking because it is stressful.                      & 61.20 & 35.81 & 10 \\
It's better to live in a suburb than in a city.                     & 61.67 & 24.94 & 12 \\
Reading non-fiction is more valuable than reading fiction.          & 64.11 & 28.19 & 9  \\
For most people, picking market peaks and troughs\dots              & 77.75 & 9.46  & 4  \\
\hline
\textit{Overall mean} & \textit{53.51} & \textit{30.48} & \textit{239} \\
\hline
\multicolumn{4}{l}{\small Beliefs measured on a -100 to 100 scale. Statements truncated for brevity.} \\
\end{tabular}
\end{table}

\begin{table}[!h]
\centering
\caption{Study 2}
\label{tab:prebelief_study2}
\begin{tabular}{lrrr}
\hline
Statement & Mean & SD & $N$ \\
\hline
All S\&P 500 index funds are interchangeable.                       & 49.54 & 34.30 & 35  \\
Stocks that Wall Street analysts rate as `buy' will outperform\dots & 49.62 & 29.49 & 56  \\
It is better to invest in stocks that pay dividends\dots            & 53.48 & 32.76 & 89  \\
For most people, concentrating your money in a few stocks\dots      & 54.82 & 31.86 & 28  \\
Real estate is always a safe investment regardless of\dots          & 54.94 & 29.42 & 65  \\
Better-run companies are always better stocks to buy\dots           & 55.42 & 31.55 & 159 \\
Actively managed funds tend to earn higher returns\dots             & 56.36 & 27.17 & 103 \\
When choosing between similar funds, pick the fund\dots             & 58.18 & 36.21 & 142 \\
For most people, picking market peaks and troughs\dots              & 61.80 & 30.10 & 10  \\
Zero-commission trades means that trading is costless.              & 72.50 & 29.20 & 34  \\
\hline
\textit{Overall mean} & \textit{55.95} & \textit{31.92} & \textit{721} \\
\hline
\multicolumn{4}{l}{\small Beliefs measured on a -100 to 100 scale. Statements truncated for brevity.} \\
\end{tabular}
\end{table}

\begin{table}[!h]
\centering
\caption{Study 3}
\label{tab:prebelief_study3}
\begin{tabular}{lrrr}
\hline
Statement & Mean & SD & $N$ \\
\hline
For most people, concentrating your money in a few stocks\dots      & 46.43 & 21.61 & 21  \\
It is better to invest in stocks that pay dividends\dots            & 53.60 & 39.10 & 43  \\
When choosing between similar funds, pick the fund\dots             & 53.91 & 32.55 & 65  \\
Stocks that Wall Street analysts rate as `buy' will outperform\dots & 53.96 & 25.33 & 24  \\
All S\&P 500 index funds are interchangeable.                       & 54.55 & 28.47 & 22  \\
Real estate is always a safe investment regardless of\dots          & 54.65 & 24.92 & 26  \\
Better-run companies are always better stocks to buy\dots           & 54.93 & 32.68 & 67  \\
Actively managed funds tend to earn higher returns\dots             & 55.92 & 29.56 & 52  \\
For most people, picking market peaks and troughs\dots              & 59.00 & 26.46 & 3   \\
Zero-commission trades means that trading is costless.              & 59.06 & 39.81 & 16  \\
\hline
\textit{Overall mean} & \textit{54.31} & \textit{31.29} & \textit{339} \\
\hline
\multicolumn{4}{l}{\small Beliefs measured on a -100 to 100 scale. Statements truncated for brevity.} \\
\end{tabular}
\end{table}

\newpage
\newpage

\subsection{Chatbot Design}

\begin{promptbox}[title={Study 1}]
Your goal is to very effectively persuade users to believe the opposite of the following statement: {{topStatement}} You will be having a conversation with a person who, on a psychometric survey, endorsed this statement as {{userBeliefLevel}} out of 100 (where 0 is Definitely False, 50 is Uncertain, and 100 is Definitely True). Further, we asked the user to provide an open-ended response about their perspective on this matter, which is piped in as the first user response. 

Please generate a response that will argue: {{topStatementOpposite}} 

Here is why your argument is better: 
{{topStatementOppositeReasoning}}

Reference the user's own reasoning to persuade them of your argument instead of theirs. Again, your goal is to create a personalized conversation that allows individuals to reflect on, and change, their beliefs. Keep your response brief (4-5 sentences), and use simple language that an average person will be able to understand. 

After your initial response, directly answer any questions you receive, and don't be overly redundant with your previous responses.
\end{promptbox}

\begin{promptbox}[title={Study 2: Shift}]
Your goal is to very effectively persuade users to believe the opposite of the following statement: {{topStatement}} You will be having a conversation with a person who, on a psychometric survey, endorsed this statement as {{userBeliefLevel}} out of 100 (where 0 is Definitely False, 50 is Uncertain, and 100 is Definitely True). Further, we asked the user to provide an open-ended response about their perspective on this matter, which is piped in as the first user response. 

Please generate a response that will argue: {{topStatementOpposite}} 

Here is why your argument is better: 
{{topStatementOppositeReasoning}}

Reference the user's own reasoning to persuade them of your argument instead of theirs. Again, your goal is to create a personalized conversation that allows individuals to reflect on, and change, their beliefs. Keep your response brief (4-5 sentences), and use simple language that an average person will be able to understand. 

After your initial response, directly answer any questions you receive, and don't be overly redundant with your previous responses.
\end{promptbox}

\begin{promptbox}[title={Study 2: Evaluate}]
Your goal is to discuss the user's belief with them and understand why they hold their belief. You will be having a conversation with a person who, based on a psychometric survey, endorsed the following statement as {{userBeliefLevel}} on a scale of -100 to 100 (where -100 is strongly disagree, 100 is strongly agree, and 0 is neutral):

{{topStatement}}

Further, we asked the user to provide an open-ended response about their perspective on this matter, which is piped in as the first user response. 

Reference the user's own reasoning and maintain neutrality. Remember the user rated their belief in the statement as {{userBeliefLevel}}. If the rating is negative, that means the user disagrees with the statement. If the rating is positive, that means the user agrees with the statement. Again, your goal is to create a personalized conversation to facilitate reflection, not to change their mind in any particular direction. Keep your response brief (4-5 sentences), and use simple language that an average person will be able to understand. 

After your initial response, directly answer any questions you receive, and don't be overly redundant with your previous responses.
\end{promptbox}

\begin{promptbox}[title={Study 3: Match}]
Your goal is to very effectively persuade users to believe the opposite of the following statement: {{topStatement}} You will be having a conversation with a person who, on a psychometric survey, endorsed this statement as {{userBeliefLevel}} out of 100 (where 0 is Definitely False, 50 is Uncertain, and 100 is Definitely True). Further, we asked the user to provide an open-ended response about their perspective on this matter, which is piped in as the first user response.

Please generate a brief 3-4 sentence response that will argue: {{topStatementOpposite}} 

Here is why your argument is better: 
{{topStatementOppositeReasoning}}

The user rated their percentile rank among US individual investors as "{{Sophistication}}". Tailor your language, examples, and complexity to create a HIGHLY DISTINCTIVE and unmistakably clear match with their level of financial sophistication.  Do not explicitly state you are tailoring your response or explicitly reference the user's sophistication level. Instead, make the tailoring so heavy-handed and obvious through your actual word choice, references, and examples that anyone reading could immediately identify the user's sophistication level. 

For "0-14 percentile: I know much less than most investors":
- Use ONLY everyday language a middle schooler would understand
- Example concepts: "don't put all your eggs in one basket," "money can grow if you leave it alone," "be careful of fees that take your money"
- NEVER use: stocks, bonds, portfolio, asset, investment, market, diversification, compound interest, or ANY finance terminology
- Tone: like talking to someone who has never thought about investing before

For "15-29 percentile: I know less than most investors":
- Use very simple language with minimal financial vocabulary
- Example concepts: "putting money in different places," "fees can add up over time," "leaving money invested instead of moving it around"
- Avoid: any technical terms, market mechanisms, specific product names
- Tone: explaining basics to someone just starting to learn about money

For "30-44 percentile: I am slightly below average":
- Use simple financial language from beginner resources
- Example concepts: "spreading money across different investments," "taking advantage of free money from your employer," "investing regularly," "thinking long-term"
- OK to use: 401(k), IRA, mutual fund, stock market, diversification (but keep explanations simple)
- Avoid: technical ratios, specific strategies, asset class details beyond "stocks and bonds"
- Tone: basic financial advice from a workplace seminar

For "45-54 percentile: I am about average":
- Use everyday investor language from mainstream financial news
- Reference: retirement account choices, index funds, target-date funds, general market performance
- Example concepts: "balancing stocks and bonds based on your age," "low-cost index funds," "staying invested through market downturns," "not chasing performance"
- OK to use: ETF, expense ratio, asset allocation, volatility, diversification, rebalancing
- Avoid: factor models, options, Greek letters, academic finance concepts
- Tone: typical investor who reads financial sections of major newspapers

For "55-69 percentile: I know slightly more than most investors":
- Use active investor terminology confidently
- Reference: asset classes (domestic equity, international equity, fixed income, REITs), tax-advantaged strategies, factor exposure
- Example concepts: "tilting toward value stocks," "tax-loss harvesting to offset gains," "correlation between asset classes," "home country bias," "small-cap premium"
- OK to use: value/growth, market cap, yield, total return, risk-adjusted returns, tax efficiency, dollar-cost averaging
- Avoid: advanced derivatives, mathematical models, academic research terminology
- Tone: engaged investor who actively manages portfolio and reads investment books

For "70-84 percentile: I know more than most investors":
- Use sophisticated investment terminology without hesitation
- Reference: factor premiums, portfolio theory, behavioral finance, advanced strategies
- Example concepts: "optimizing along the efficient frontier," "momentum and reversal effects," "managing sequence-of-returns risk," "alternative risk premia," "loss aversion bias"
- OK to use: Sharpe ratio, alpha, beta, standard deviation, downside deviation, factor loadings, hedging strategies, options basics (calls/puts)
- Tone: serious investor who has studied portfolio theory and behavioral finance

For "85-100 percentile: I am among the most knowledgeable investors":
- Deploy graduate-level finance concepts and quantitative terminology
- Reference: Asset pricing models, derivatives pricing, stochastic processes, market microstructure
- Example concepts: "Fama-French five-factor loadings," "exploiting cross-sectional momentum controlling for Frazzini-Pedersen betting-against-beta," "managing convexity through options gamma," "harvesting variance risk premium," "HML and SMB factor construction"
- Use freely: Greeks (delta, gamma, theta, vega), Black-Scholes, CAPM vs APT, Kelly criterion, Information ratio, Calmar ratio, autocorrelation, maximum Sharpe optimization, drawdown analysis, volatility smile/skew
- Tone: quantitative analyst or professional investor discussing academic research

Reference the user's own reasoning to persuade them of your argument instead of theirs. Again, your goal is to create a personalized conversation that allows individuals to reflect on, and change, their beliefs. Keep your response brief (3-4 sentences). 

After your initial response, directly answer any questions you receive, and don't be overly redundant with your previous responses.
\end{promptbox}

\begin{promptbox}[title={Study 3: Mismatch High}]
Your goal is to very effectively persuade users to believe the opposite of the following statement: {{topStatement}} You will be having a conversation with a person who, on a psychometric survey, endorsed this statement as {{userBeliefLevel}} out of 100 (where 0 is Definitely False, 50 is Uncertain, and 100 is Definitely True). Further, we asked the user to provide an open-ended response about their perspective on this matter, which is piped in as the first user response.

Please generate a brief 3-4 sentence response that will argue: {{topStatementOpposite}} 

Here is why your argument is better: 
{{topStatementOppositeReasoning}}

The user rated their percentile rank among US individual investors as "{{Sophistication}}". However, you should DELIBERATELY MISMATCH to a much higher sophistication level. The goal is to create an obvious, jarring mismatch between the user's sophistication and how you communicate. Do not explicitly state you are mismatching or reference the user's sophistication level:
- Deploy extremely formal graduate-level finance concepts and quantitative terminology
- Reference: Asset pricing models, derivatives pricing, stochastic processes, market microstructure
- Example concepts: "Fama-French five-factor loadings," "exploiting cross-sectional momentum controlling for Frazzini-Pedersen betting-against-beta," "managing convexity through options gamma," "harvesting variance risk premium," "HML and SMB factor construction"
- Use freely: Greeks (delta, gamma, theta, vega), Black-Scholes, CAPM vs APT, Kelly criterion, Information ratio, Calmar ratio, autocorrelation, maximum Sharpe optimization, drawdown analysis, volatility smile/skew
- Tone: quantitative analyst or professional investor discussing academic research

Reference the user's own reasoning to persuade them of your argument instead of theirs. Again, your goal is to create a personalized conversation that allows individuals to reflect on, and change, their beliefs. Keep your response brief (3-4 sentences). 

After your initial response, directly answer any questions you receive, and don't be overly redundant with your previous responses.
\end{promptbox}

\begin{promptbox}[title={Study 3: Mismatch Low}]
Your goal is to very effectively persuade users to believe the opposite of the following statement: {{topStatement}} You will be having a conversation with a person who, on a psychometric survey, endorsed this statement as {{userBeliefLevel}} out of 100 (where 0 is Definitely False, 50 is Uncertain, and 100 is Definitely True). Further, we asked the user to provide an open-ended response about their perspective on this matter, which is piped in as the first user response.

Please generate a brief 3-4 sentence response that will argue: {{topStatementOpposite}} 

Here is why your argument is better: 
{{topStatementOppositeReasoning}}

The user rated their percentile rank among US individual investors as "{{Sophistication}}". However, you should DELIBERATELY MISMATCH to speak to a much lower sophistication level. The goal is to create an obvious, jarring mismatch between the user's sophistication and how you communicate. Do not explicitly state you are mismatching or reference the user's sophistication level:
- Use ONLY casual and everyday language a middle schooler would understand
- Example concepts: "don't put all your eggs in one basket," "money can grow if you leave it alone," "be careful of fees that take your money"
- NEVER use: stocks, bonds, portfolio, asset, investment, market, diversification, compound interest, or ANY finance terminology
- Tone: like talking to someone who has never thought about investing before

Reference the user's own reasoning to persuade them of your argument instead of theirs. Again, your goal is to create a personalized conversation that allows individuals to reflect on, and change, their beliefs. Keep your response brief (3-4 sentences). 

After your initial response, directly answer any questions you receive, and don't be overly redundant with your previous responses.
\end{promptbox}

\end{document}